\documentclass{ws-procs10x7}
\usepackage{balance}
\usepackage{cite,./mcite}

\def\kt{\ensuremath{k_\perp}}

\newcommand{\as}{{\alpha}_\mathrm{s}}

\newcommand{\Pmax}{\bar{q}}

\newcolumntype{d}[1]{D{.}{.}{#1}}

\makeindex
\begin{document}

\title{Towards precision determination of unintegrated Parton Density Functions}

\author{H. Jung}

\address{DESY, FRG\\E-mail: hannes.jung@desy.de}

\author{A.V. Kotikov }
\address{BLTHPH, JINR, Dubna,  Russia \\E-mail: kotikov@theor.jinr.ru}
\author{ A.V. Lipatov \and N.P. Zotov }
\address{SINP, Lomonosov Moscow State University,  Russia 
\\E-mail: zotov@theory.sinp.msu.ru, lipatov@theory.sinp.msu.ru}


\twocolumn[\maketitle\abstract{
First attempts are described to determine the unintegrated Parton Density
Function of the gluon from a fit to measurements of the structure function
$F_2(x,Q^2)$ and also $F_2^c(x,Q^2)$ measured at HERA. Reasonable descriptions
of both structure functions are obtained, but the gluon densities determined are
different.
}
\keywords{pQCD; unintegrated parton densities; $\kt$-factorization.}
]

\section{Introduction}
Unintegrated 
parton density functions (uPDFs) 
are best suited to study details of the
hadronic final state in high energy $ep$ and also in $pp$ collisions (for a
review see
\cite{jung-hq-2001,jung-mpla2003,smallx_2001,smallx_2002,
smallx_2004,jung-dis04,jung-collins-2005}). 
In general, the production 
cross section for jets, heavy quarks or gauge bosons
can be
written as a convolution of the uPDF ${\cal A}(x,\kt^2,\Pmax)$ with the 
partonic off-shell cross section ${\hat\sigma}(x_i, \kt^2)$, with 
$x_i, \kt$ being the longitudinal momentum fraction and the transverse momentum
of the interacting parton $i$ and $\Pmax $ being the factorization scale. For
example the cross section for  $ep \to \mbox{jets} + X $ can be written as:
\begin{eqnarray}
\frac{d \sigma^{jets}}{d E_T d \eta} &=&
\sum_i \int\int\int dx_i\; dQ^2 d\dots \nonumber \\[4pt]
&&{} 
\left[ d\kt^2 x_i{\cal A}(x_i,\kt^2,\Pmax)\right] {\hat\sigma} (x_i, \kt^2)
\nonumber \end{eqnarray}
At high energies, the gluon density
 is dominating for many processes, therefore 
here only the gluon uPDF is considered. It has already been
shown in \cite{heralhc2006a,*heralhc-hq}, that the predictions 
of the total cross section as well as  
differential distributions for heavy quark production at HERA
and the LHC agree well in general with those coming from  fixed NLO 
calculations. However, the details
depend crucially on a precise knowledge of the uPDF. Therefore precision fits to
inclusive and exclusive measurements have to be performed to determine precisely
the free parameters of the uPDF: the starting distribution
function at a low scale $\Pmax_0 \sim 1$ GeV as well as parameters connected
with $\as$ and details of the splitting functions for the perturbative evolution. 

The previously 
available uPDFs (a overview is given in \cite{smallx_2001,smallx_2002,
smallx_2004})
were only adjusted to describe $F_2$, but no real fit of the
parameters of the starting distribution was performed, 
neither the experimental
uncertainties were treated as is done in global QCD analyses 
in the collinear approach
(for example from CTEQ
or MRST
).

Here attempts to determine the uPDF of the gluon from pQCD fits using the CCFM
evolution equation \cite{CCFMa,*CCFMb,*CCFMc,*CCFMd}.
to the inclusive
structure function $F_2(x,Q^2)$ as well as $F_2^c(x,Q^2)$, as measured by the
HERA experiments, are described.

\section{The method}
The
unintegrated gluon density is determined by a convolution of the
non-perturbative starting distribution ${\cal A}_0 (x)$ and
the CCFM evolution denoted by 
${\cal \tilde A}\left(x,\kt,\Pmax\right)$: 
\begin{eqnarray}
x {\cal A}(x,\kt,\Pmax) &= &\int dx' 
{{\cal A}_0 (x') }\cdot \frac{x}{x'}\nonumber \\[4pt]
&&{}{ {\cal \tilde A}\left(\frac{x}{x'},\kt,\Pmax\right) }\nonumber 
\end{eqnarray}
In the perturbative evolution  the  gluon splitting function $P_{gg}$ 
including non-singular terms 
(as described in detail in \cite{jung-dis02,*jung-dis03}) is applied.

The distribution ${\cal A}_0$ is parameterized at the starting scale $\Pmax_0$
by:
\begin{equation}
x{\cal A}_0(x) = N x^{-B_g} \cdot (1 -x)^{C_g}\left( 1 -D_g x\right)\nonumber 
\end{equation}
The parameters $N_g, B_g, C_g, D_g$ of ${\cal A}_0$ are determined in the 
fit, which is based on the MINUIT package~\cite{minuit} and the extension by
\cite{Pumplin:2000vx}, such to minimize the $\chi^2$ defined by:
\begin{equation}
\chi^2 = 
\sum_i \left( \frac{\left(T - D  \right)^2}
{\sigma_i^{2\;stat} + \sigma_i^{2\;uncor}}\right) \nonumber 
\end{equation}
with $T $ being the theory value and $D$ the measurement with 
the corresponding statistical and uncorrelated systematic uncertainty.

\section{The structure function $F_2(x,Q^2)$}
\begin{figure}[htb]
\vspace*{-0.1cm}
\includegraphics[width=0.5\textwidth]{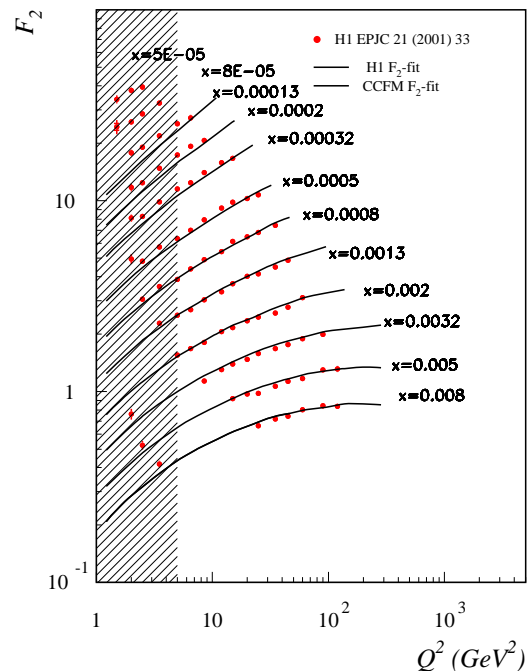}
\caption{The structure function $F_2(x,Q^2)$ as a function of $Q^2$ for
different values of $x$. The data points are from \protect\cite{h1_f2-00181}.
The shaded area shows the region which is not used in the fit. }
\label{F2}
\end{figure}

\begin{figure}[h]
\includegraphics[width=0.5\textwidth]{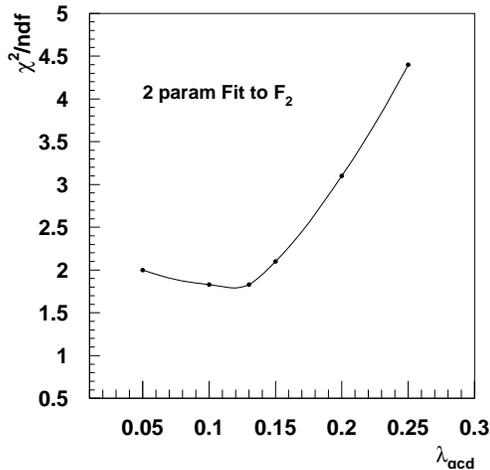}
\caption{$\chi^2/ndf$ dependence of the $F_2$ fit on the choice of
$\Lambda_{qcd}$ }
\label{lambda}
\end{figure}

The measurement of $F_2(x,Q^2)$ \cite{h1_f2-00181} is used in the range $x < 0.005$
and $Q^2 > 5$ GeV$^2$ to determine the uPDF. 
The parameters
of ${\cal A}_0$ were investigated separately, and it was confirmed, that the
data do not constrain the parameters $C_g $ and $D_g$, which were therefore fixed
to $C_g=4$ and $D_g =0 $. The starting distribution ${\cal A}_0$ was
parameterized at $\Pmax_0 = 1.2 $ GeV.
The running coupling $\as(\mu)$ was used in the 1-loop
approximation in the region $\mu > \Pmax_0$, and was kept fixed at
$\as(\Pmax_0)$ for $\mu < \Pmax_0$. A acceptable fit to the measured $F_2$
values was obtained with  $\chi^2/ndf = 111.8/61 = 1.83$ 
using only statistical
and uncorrelated systematical uncertainties 
(compare to $\chi^2/ndf \sim 1.5$ in
the collinear approach at NLO). 
In Fig.~\ref{F2} the measurement is compared to the prediction of the structure
function $F_2(x,Q^2)$ as obtained from the fit. 
Since in the cross section a
product of the gluon density and $\as$ enters (as well as $\as$ enters in the
pQCD evolution) also the  dependence on the choice of
$\Lambda_{qcd}$ was investigated (Fig.~\ref{lambda}). 
A clear preference for 
$\Lambda_{qcd}^{(4)} \sim 0.13$ is observed, which corresponds to
$\as(m_{Z_0})=0.118$ in the 1-loop approximation.

\section{The structure function $F_2^c(x,Q^2)$ and the uPDF}
\begin{figure}[b]
\vspace*{-0.1cm}
\includegraphics[width=0.5\textwidth]{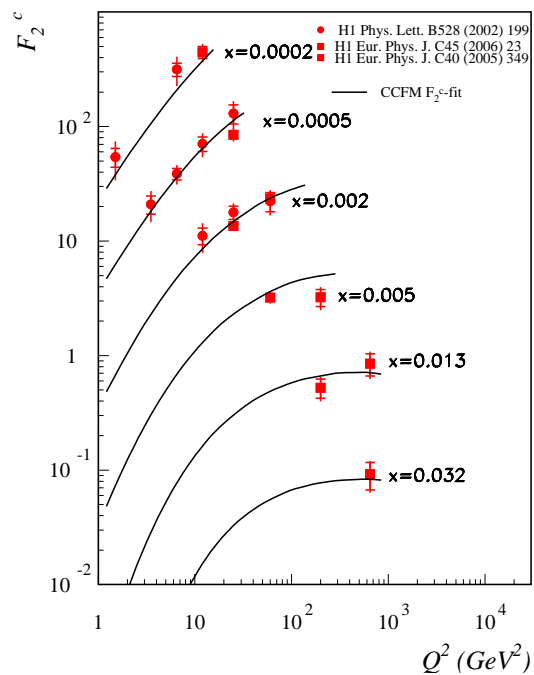}
\caption{The structure function $F_2^c(x,Q^2)$ as a function of $Q^2$ for
different values of $x$. The data points are from 
\protect\cite{h1_f2c_dstar,*h1_f2c_vtx_lowq2,*h1_f2c_vtx_higq2}. }
\label{F2c}
\vspace*{-0.1cm}
\end{figure}

The structure function $F_2^c(x,Q^2)$ is directly sensitive to the gluon
density. The measurements of
\cite{h1_f2c_dstar,*h1_f2c_vtx_lowq2,*h1_f2c_vtx_higq2} were used to determine
the uPDF in the range $Q^2 \geq 1.5 $ GeV$^2$. A acceptable fit to 
the measured $F_2^c$
values was obtained with  $\chi^2/ndf = 18.8/ 20= 0.94$ using  statistical
and systematic uncertainties. In Fig.~\ref{F2c} the measurement 
is compared to the prediction of the structure
function $F_2^c(x,Q^2)$ as obtained from the fit.
\par
\begin{figure}[ht]
\vspace*{-0.1cm}
\includegraphics[width=0.5\textwidth]{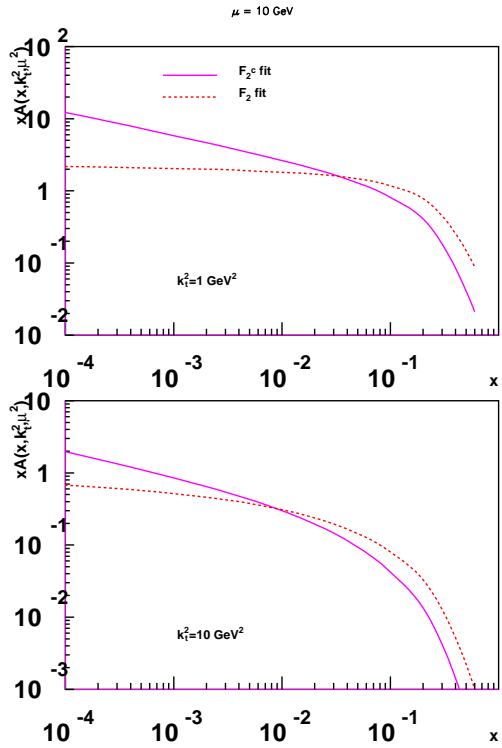}
\caption{The unintegrated gluon density as obtained from the fits to 
$F_2(x,Q^2)$ and $F_2^c(x,Q^2)$ as a function of $x$ for different values of
$\kt$. The gluon density is evolved to a scale $\Pmax = 10 $ GeV.
}
\label{updf}
\vspace*{-0.1cm}
\end{figure}

Since the fits to  $F_2(x,Q^2)$ and $F_2^c (x,Q^2)$ were performed independently,
it is interesting to compare the resulting unintegrated gluon distributions
(Fig.~\ref{updf}). The small $x$ behavior of the 
gluon distribution is very
different. It has been explictly checked, that 
the measured $F_2(x,Q^2)$
cannot be described with the uPDF obtained from the fit to $F_2^c(x,Q^2)$ and
vice versa.

\section{The structure function $F_L^c(x,Q^2)$ }
\begin{figure}[ht]
\vspace*{-0.1cm}
\includegraphics[width=0.5\textwidth]{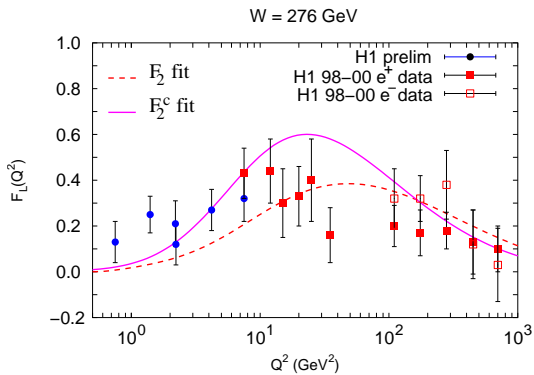}
\caption{The longitudinal structure function \newline $F_L(x,Q^2)$  at fixed $W$ as
measured by \protect\cite{Newman:2003vh,*Lobodzinska:2003yd}. 
The solid curve is solid curve is the prediction using the fit to $F_2$ whereas
the dashed curve uses the uPDF obtained from a fit to $F_2^c$.
}
\label{FL}
\vspace*{-0.1cm}
\end{figure}

Using the different uPDFs described in the previous sections we calculate the
longitudinal structure function $F_L(x,Q^2)$ 
at fixed $W$ in the framework of 
\cite{Kotikov:2004uf,*Kotikov:2005np}. The predictions are compared with the measurements 
at fixed $W$ of
H1~\cite{Newman:2003vh,*Lobodzinska:2003yd} in Fig.~\ref{FL}. The general trend
is nicely described, however the predicted shape of $F_L$ at medium $Q^2$ is
different and precise measurements would be very welcome. 
\section*{Acknowledgments}
Many thanks go the 
E. Perez for the support with the fit program and the experimental error
treatment. Thanks also to the organizers of this very interesting conference.

\providecommand{\etal}{et al.\xspace}
\providecommand{\href}[2]{#2}
\providecommand{\coll}{Coll.}
\catcode`\@=11
\def\@bibitem#1{%
\ifmc@bstsupport
  \mc@iftail{#1}%
    {;\newline\ignorespaces}%
    {\ifmc@first\else.\fi\orig@bibitem{#1}}
  \mc@firstfalse
\else
  \mc@iftail{#1}%
    {\ignorespaces}%
    {\orig@bibitem{#1}}%
\fi}%
\catcode`\@=12

\end{document}